\documentclass{ws-procs9x6-cpt25}

\usepackage{graphicx}
\usepackage{amsmath}
\usepackage{slashed}
\usepackage{multirow}
\usepackage{subfigure}
\usepackage{xspace}
\usepackage{relsize}

\usepackage{xcolor} 

\usepackage{pagecolor}
\pagecolor{white}

\def\etal{{\it et al.}}
 
\def\babar{\mbox{\slshape B\kern-0.1em{\smaller A}\kern-0.1em
    B\kern-0.1em{\smaller A\kern-0.2em R}}\xspace}

\def\nn#1#2#3{\hskip -6pt{}^{\phantom{#3}#2}_{\phantom{#2}#3}#1}
\def\degree{^\circ}
\newcommand{\lsim}{\mathrel{\rlap{\lower4pt\hbox{\hskip1pt$\sim$}}
   \raise1pt\hbox{$<$}}}
      
\begin{document}

\title{Flavor-changing signatures of spacetime-symmetry violations}

\author{N.\ Sherrill}

\address{Institut f\"ur Theoretische Physik, Leibniz Universit\"at Hannover,\\
Appelstraße 2, Hannover, 30167, Germany}

\begin{abstract}
This proceedings contribution discusses Lorentz and CPT violation 
in the context of charged-lepton flavor-changing processes. 
Particular emphasis is placed on
coherent muon-to-electron conversion in muonic atoms.
Data collected by the SINDRUM~II experiment 
at the Paul Scherrer Institute is used to extract
constraints on mass dimension $d=5$ electromagnetic 
and $d=6$ quark-lepton effects.
The Mu2e experiment at Fermilab
and COMET experiment at the 
Japan Proton Accelerator Complex are projected to
improve constraints by up to two orders in magnitude.
\end{abstract}

\bodymatter

\section{Introduction}
Charged-lepton flavor-changing (CLFC) processes initiated by
Standard Model (SM) interactions in conjunction with
neutrino oscillations yield practically unobservable effects.~\cite{smdecays} Experiments in search of CLFC processes thus provide clean probes of physics beyond the SM.~\cite{cs18,bh94,bc13} 
This contribution examines CLFC processes initiated
by small violations of Lorentz and CPT invariance, with an emphasis on neutrinoless and coherent muon-to-electron conversion in muonic atoms.~\cite{Kostelecky:2025imu}

A model-independent description of Lorentz violation (LV) and other fundamental symmetries is provided by the Standard-Model Extension (SME) effective field theory.~\cite{ck97,ak04,kl19,review} The SME contains operators describing CLFC interactions,
of which only a subset involving muon decays has been investigated.~\cite{tables, ck97,cg99,yi03,gh20,cks20,hp11,nowt16,lmt16,msf17,gmn20,kps22}
Using existing data collected by the SINDRUM~II experiment 
at the Paul Scherrer Institute,~\cite{sindrum} we 
revisit constraints on mass dimension $d=5$ operators first extracted from electromagnetic decays~\cite{kps22} and deduce several first constraints on
$d=6$ quark-lepton operators. 

One golden CLFC channel 
is the coherent conversion $\mu + N \to e + N$
of a muon $\mu$ into an electron $e$
in the presence of a nucleus $N$. Experiments 
searching for this process involve
a proton beam that strikes a production target,
generating pions and other hadrons that decay to muons among other products. The muons are further directed onto a stopping target, forming muonic atoms that relax to the 1S ground state.
The signal for a neutrinoless muon-electron conversion 
is a monoenergetic electron ejected from a muonic atom with energy
$E_e^{\rm conv} = m_\mu - E_{\rm bind} - E_{\rm recoil}$,
where $m_\mu$
is the muon mass, 
$E_{\rm bind}$
is the muon binding energy,
and $E_{\rm recoil}$ 
is the nuclear recoil energy.
This process can be characterized experimentally
by the dimensionless ratio 
\begin{equation}
R_{\mu e} = 
\frac{\omega_{\rm conv}}{\omega_{\rm capt}}, 
\label{R}
\end{equation}
where $\omega_{\rm conv}$ and $\omega_{\rm capt}$ 
are the muon conversion and capture rates, respectively. 
The dominant contribution to $\omega_{\rm conv}$
is the coherent conversion
$\mu^- + \nn NAZ \rightarrow e^- + \nn NAZ$,
where the nucleus $\nn NAZ$ of the muonic atom 
has atomic number $A$ and charge $Z$.~\cite{cohconv}

To date, SINDRUM~II
has achieved the most stringent constraint,
$R_{\mu e} < 7\times 10^{-13}$ at the $90\%$ confidence level (CL),
using~$\nn{\rm Au}{197}{79}$ as the target ($\omega^{\rm Au}_{\rm capt} \simeq 13.07$~MHz~\cite{ts87}).~\cite{sindrum} 
Upcoming searches using an~$\nn{\rm Al}{27}{13}$ target ($\omega^{\rm Al}_{\rm capt} \simeq 0.7054$~MHz~\cite{ts87})
include 
the Muon-to-Electron Conversion (Mu2e) experiment 
at Fermilab,
which anticipates achieving $R_{\mu e} \simeq 6.2\times 10^{-16}$,~\cite{mu2e}
and the Coherent Muon to Electron Transition (COMET) 
experiment
at the Japan Proton Accelerator Complex (J-PARC),
which expects to reach $R_{\mu e} \simeq 7\times 10^{-15}$.~\cite{comet}
More distant future upgrades include
phase two of Mu2e,~\cite{mu2e2}
and phase two of COMET~\cite{comet2} 
and other experiments at J-PARC,
which expect to reach $R_{\mu e} \simeq 10^{-18}$.~\cite{prismprime}

No observable CLFC effects in muon decay 
involve SME terms of mass dimension $d=3$ or $4$.~\cite{kps22}
The leading contributions instead arise 
via $d=5$ electromagnetic interactions 
and $d=6$ (4-point) quark-lepton interactions
rather than from propagator terms.~\cite{muon34}
Since the 4-point interactions 
involve the quarks in the nucleus,
the corresponding CLFC effects
are uniquely accessed in the coherent conversion $\mu + N \to e + N$.
We report the first constraints on these latter effects.~\cite{Kostelecky:2025imu} 

In the following sections, we calculate the leading SME contributions to the conversion rate $\omega_{\rm conv}$ for muonic gold and aluminum targets. Constraints are extracted from the SINDRUM~II result,~\cite{sindrum} and projected constraints are estimated for the upcoming Mu2e and COMET experiments.

\section{Setup}
The muon and electron wavefunctions are written as 
\begin{equation}
\psi_{\kappa,s} (r, \theta, \phi) = 
{g(r) \chi_{\kappa,s} (\theta,\phi) 
\choose if(r) \chi_{-\kappa,s}(\theta,\phi)},
\end{equation}
where $\kappa = \mp(J+1/2)$ 
and the eigenspinors $\chi_{\kappa, s}$ satisfy 
$(\boldsymbol{\sigma} \cdot \boldsymbol{l}+I)\chi_{\kappa,s}
=-\kappa\chi_{\kappa,s}$ 
and $J_z\chi_{\kappa,s}=s\chi_{\kappa,s}$, 
with normalization 
$\int \,d\Omega~\chi_{\kappa^\prime,s^\prime}
^\dagger\chi_{\kappa,s}
=\delta_{\mu^\prime\mu}\delta_{\kappa^\prime\kappa}$. 
The 1S muon wavefunction has $\kappa = -1$, 
and neglecting the electron mass implies $g^+ = if^-$ and $f^+ = -ig^-$.
The radial wavefunctions are obtained by numerically 
solving the Dirac equation 
\begin{equation}
\frac{d}{dr}
\begin{pmatrix} 
u_1 \\ u_2
\end{pmatrix} = 
\begin{pmatrix}
-\kappa/r & W-V(r)+m\\
-(W-V(r)-m) & \kappa/r
\end{pmatrix}
\begin{pmatrix} u_1 \\ u_2\end{pmatrix}
\end{equation}
for $u_1(r)=rg(r)$ and $u_2(r)=rf(r)$.
Here, 
$W$ is the lepton energy
and $V(r)$ is the approximately static and spherically symmetric nuclear electric potential.~\cite{nuclsph}
The potential is obtained 
from $\rho(r)$,
which is normalized as $\int_{0}^{\infty} 4\pi\rho(r)r^2 \,dr = Z$.
The form of $\rho(r)$ is taken as 
a two-parameter Fermi model for$\nn{\rm Au}{197}{79}$
and as a Fourier-Bessel expansion for$\nn{\rm Al}{27}{13}$.~\cite{vjv87,kko02}

The conversion rate for $d=5$ electromagnetic interactions may be written as
\begin{equation}
\omega_{\rm conv}
= \tfrac{1}{2} \hskip -4pt 
\sum_{s=\pm\frac{1}{2}} 
\sum_{\kappa = \pm 1} 
\sum_{s^\prime=\pm\frac{1}{2}} 
\left| \int d^3x 
F_{\alpha\beta} 
\overline{\psi}{}_{\kappa,s'}^{(e)} 
\mathcal{O}^{\alpha\beta} \psi_{s}^{(\mu)} 
\right|^2,
\label{d5rate}
\end{equation}
where 
$F_{\alpha\beta}$ is the electromagnetic field strength, 
$\psi_{\kappa,s'}^{(e)}$ 
is a continuum electron wavefunction,
and 
$\psi_{s}^{(\mu)}$ 
is the 1S bound-state muon wavefunction.
The matrix ${\cal O}^{\alpha\beta}$
contains the set
$\{(m_{F}^{(5)})^{\alpha\beta}_{\mu e} $,
$i(m_{5F}^{(5)})^{\alpha\beta}_{\mu e}\gamma_5 $,
$(a_{F}^{(5)})^{\lambda\alpha\beta}_{\mu e}\gamma_\lambda $,
$(b_{F}^{(5)})^{\lambda\alpha\beta}_{\mu e}\gamma_5\gamma_\lambda $,
$\tfrac{1}{2}
(H_{F}^{(5)})^{\kappa\lambda\alpha\beta}_{\mu e}\sigma_{\kappa\lambda} \}$.~\cite{kl19,kps22}
These modify interactions rather than propagators,
so standard perturbative quantization methods hold~\cite{kl01}
and~\eqref{d5rate}
can be derived using tree-level Feynman rules.

The conversion rate for the $d=6$ quark-lepton interactions may be written as
\begin{equation}
\omega_{\rm conv} 
= \tfrac{1}{2} \hskip -4pt 
\sum_{s,s',\kappa}  
\left| \int d^3x 
\left(
\alpha
\overline{\psi}{}_{\kappa,s'}^{(e)} 
\mathcal{K} \psi_{s}^{(\mu)}
+
\beta
\overline{\psi}{}_{\kappa,s'}^{(e)} 
\mathcal{K}_0 \psi_{s}^{(\mu)}
\right)
\right|^2,
\label{d6rate}
\end{equation}
where $\alpha$ and $\beta$ are conventional nuclear matrix elements
given by
$\alpha = \langle N| \overline{\psi}{}^{(q)} \psi^{(q)} |N\rangle$
with $q$ summed over the quark flavors $q = u,d,s$
and 
$\beta = \langle N| \overline{\psi}{}^{(q)} \gamma_0 \psi^{(q)}|N \rangle$
with $q$ summed over $q = u,d$.~\cite{kks01}
Other possible nuclear matrix elements vanish 
in coherent conversion.~\cite{kko02}
The matrices ${\cal K}$ and ${\cal K}_0$ 
incorporate the $d=6$ SME coefficients,~\cite{kl19}
where ${\cal K}$ contains the set 
$\{(k_{SV}^{(6)})^{\lambda}_{qqe\mu}\gamma_\lambda$,
$(k_{SA}^{(6)})^{\lambda}_{qqe\mu}\gamma_5\gamma_\lambda$,
$\tfrac{1}{2}
(k_{ST}^{(6)})^{\kappa\lambda}_{qqe\mu}\sigma_{\kappa\lambda}\}$
for $q = u,d,s$
and 
${\cal K}_0$ 
contains the set
$\{(k_{VS}^{(6)})^{t}_{qqe\mu}$,
$i(k_{VP}^{(6)})^{t}_{qqe\mu}\gamma_5$,
$(k_{VV}^{(6)})^{t\lambda}_{qqe\mu}\gamma_\lambda$,
$(k_{VA}^{(6)})^{t\lambda}_{qqe\mu}\gamma_5\gamma_\lambda$,
$\tfrac{1}{2}
(k_{VT}^{(6)})^{t\kappa\lambda}_{qqe\mu}\sigma_{\kappa\lambda}\}$
for $q = u,d$.

\section{Experiments and constraints}
For the experiments considered here,
the cartesian SME coefficients 
appearing in ${\cal O}^{\alpha\beta}, {\cal K}$, and ${\cal K}_0$
can be taken to be independent of time and location.~\cite{ak04}
The coefficients change under observer Lorentz transformations,
implying measurements of their values
must be provided in a specified inertial frame.
The canonical choice is the Sun-centered frame (SCF)
with cartesian coordinates $(T,X,Y,Z)$. 
The time origin $T=0$ is defined as the 2000 vernal equinox,
the $X$ axis is chosen to point from the Earth to the Sun at $T=0$,
and the $Z$ axis is parallel to the Earth's rotation axis.~\cite{sunframe}
To establish the transformation from the SCF 
to the laboratory detector frame (DF), 
it is also useful to introduce a standard Earth-based laboratory frame (LF)
with cartesian coordinates $(x,y,z)$
and the $x$ ($y$) axis pointing to local south (east).

Laboratory frames are typically noninertial
due to the rotation of the Earth at the sidereal frequency
${\omega_\oplus}\simeq 2\pi /(23\,{\rm h}~56\,\min)$.
At leading order, the transformation from the SCF to the LF 
is a rotation $\mathcal{R}(\chi, \omega_\oplus T_\oplus)$
that depends on the laboratory colatitude~$\chi$ 
and harmonics~\cite{ak98} of the laboratory sidereal time
$T_\oplus \equiv T - T_0$.
The time $T_\oplus$ in the LF 
is shifted by an experiment-dependent amount $T_0$
relative to the time $T$ in the SCF,
arising from the laboratory longitude $\lambda$ 
and other effects.~\cite{offset}
The explicit form of the rotation $\mathcal{R}(\chi, \omega_\oplus T_\oplus)$ 
is given as Eq.~(7) of Ref.~\cite{kps22}.
The DF typically differs from the standard LF,
so we must perform an additional transformation 
$\mathcal{R_{\text{detector}}}(\psi)$
given in Eq.~(6) of Ref.~\cite{kps22}
that involves the angle $\psi$ of the laboratory $z$-axis
measured north of east along the direction of the beamline. The net transformation from the SCF to the DF is thus $\mathcal{R_{\text{total}}} = 
\mathcal{R_{\text{detector}}}(\psi) 
\mathcal{R}(\chi, \omega_\oplus T_\oplus)$. 
Applying the transformation  $\mathcal{R_{\text{total}}}$
yields $\omega_{\rm conv} = \omega_{\rm conv}(\psi, \chi, \omega_\oplus T_\oplus)$, and thus  $R_{\mu e} = R_{\mu e} (\psi, \chi, \omega_\oplus T_\oplus)$, in terms of the detector orientation, detector location, sidereal harmonics, and the SME coefficients in the SCF. 
The quantities $\chi$, $\lambda$, $T_0$, and $\psi$
for the experiments considered here are listed 
in Table~\ref{exptprops}.

\begin{table*}
\centering
\setlength{\tabcolsep}{5pt}
\begin{tabular}{c | c c c c}
\hline\hline
Experiment $	$&$	\chi	$&$	\lambda	$&$	T_0	$&$	\psi$ \\	\hline
SINDRUM II $	$&$	42.5\degree	$&$	8.2\degree	$&$	3.86\;\rm{h} 	$&$	242\degree$\\
COMET $	$&$	53.6\degree	$&$	140.6\degree	$&$	-4.94\;\rm{h} 	$&$	188\degree$ \\
Mu2e $	$&$	48.2\degree	$&$	-88.2\degree	$&$	10.27\;\rm{h} 	$&$	122\degree$\\
\hline\hline 
\end{tabular}
\caption{Detector-frame parameters.}
\label{exptprops}
\end{table*}

The spherical symmetry of the nuclear charge distribution
implies contributions to Eq.~\eqref{d5rate} arise only from the field strength components
$F_{tr} = -F_{rt} = E_r(r)$,
where the index $t$ represents the time $T_\oplus$. 
As ${\cal O}^{tr}$ is a spatial vector,
the match between polar and cartesian coordinates in the DF
can be accomplished by 
$\mathcal{O}^{tr} = 
\mathcal{O}^{tx} \sin{\theta}\cos{\phi}
+\mathcal{O}^{ty} \sin{\theta}\sin{\phi}
+\mathcal{O}^{tz} \cos{\theta}$,
as usual. Working in DF spherical-polar coordinates with the $z$ axis directed along the muon beam direction,
the azimuthal integral can then be evaluated over the full range $\phi\in[0,2\pi)$. 
In contrast, the polar angle $\theta$ is limited by the detector acceptance
to a range $|\cos\theta|\lsim c$. Integration over $\theta$ results in two geometrical factors $w_{xy} = -c^3/6+c/2$ and $w_z = c^3/3$.  The approximate values of the acceptances $c$ 
are listed in Table~\ref{exptprops2}.

\begin{table*}
\centering
\setlength{\tabcolsep}{5pt}
\begin{tabular}{c | c c c c c}
\hline\hline
Experiment $	$&$	c  $&$ \widetilde{I}_1	 $&$	\widetilde{I}_2	$&$	\widetilde{I}_3	$&$	\widetilde{I}_4	$\\	\hline
SINDRUM II $	$&$	0.44	$&$	-0.053	$&$	0.070	$&$	0.040	$&$	0.18	$\\
COMET $	$&$	0.34	$&$	-0.012	$&$	0.012	$&$	7.6\times 10^{-4}	$&$	0.041	$\\
Mu2e $  $&$   0.50	$&$	-0.012	$&$	0.012	$&$	7.6\times 10^{-4}	$&$	0.041	$\\
\hline\hline 
\end{tabular}
\caption{Detector acceptances and normalized radial integrals.}
\label{exptprops2}
\end{table*}
The radial part of Eq.~\eqref{d5rate} gives two integrals unique to each experiment,
\begin{align}
I_1 =& \int \,dr~r^2 E(r) \left(f_e^-g_\mu^-+g_e^-f_\mu^-\right),
\nonumber\\
I_2 =& \int \,dr~r^2 E(r) \left(g_e^-f_\mu^--f_e^-g_\mu^-\right).
\label{integrals}
\end{align}
Table \ref{exptprops2} displays numerical values of the quantities 
$\widetilde{I}_1 = I_1/m_\mu^{3/2}$ and $\widetilde{I}_2 = I_2/m_\mu^{3/2}$.
Combining these results gives the conversion rate $\omega_{\rm conv}$
expressed using SME coefficients defined in the DF.

Time averaging can be performed for data collected over extended intervals,
leaving only the time-independent contribution 
$\overline{\omega}_{\rm conv}$.
Following accepted procedure, an experimental measurement of $\overline{\omega}_{\rm conv}$
can thereby be translated into constraints 
on the components of the SME coefficients
taken one at a time.
The SINDRUM~II result $R_{\mu e} < 7\times 10^{-13}$
was obtained using a primary dataset taken 
over 81 days.~\cite{sindrum}
It can therefore be intepreted as a limit 
on the time-averaged conversion rate 
$\overline{\omega}_{\rm conv} = 
\overline{\omega}_{\rm conv} (\psi, \chi)$
for the corresponding $\psi$ and $\chi$ values
given in Table~\ref{exptprops}.

With $R_{\mu e}, \omega_{\rm capt}$, and $\overline{\omega}_{\rm conv}$, 
constraints on the $d=5$ 
SME coefficient components are extracted and presented in Table~\ref{resultsd5}. 
\renewcommand\arraystretch{0.8}
\begin{table}
\centering
\setlength{\tabcolsep}{2pt}
\begin{tabular}{c c c c c}
\hline\hline								
Coefficients	&	SINDRUM~II	&	COMET &	Mu2e	\\	
\hline								
$ |(m_{F}^{(5)})^{TJ}_{\mu e}|, |(m_{5F}^{(5)})^{TJ}_{\mu e}|	$&$	< 8	$&$	< 1	$&$	< 0.2	$ \\
$ |(m_{F}^{(5)})^{TZ}_{\mu e}|, |(m_{5F}^{(5)})^{TZ}_{\mu e}|	$&$	< 8	$&$	< 0.9	$&$	< 0.2	$ \\ [4pt]
$ |(a_{F}^{(5)})^{TTJ}_{\mu e}|, |(b_{F}^{(5)})^{TTJ}_{\mu e}|	$&$	< 6	$&$	< 1	$&$	< 0.2	$ \\
$ |(a_{F}^{(5)})^{TTZ}_{\mu e}|, |(b_{F}^{(5)})^{TTZ}_{\mu e}|	$&$	< 6	$&$	< 0.9	$&$	< 0.2	$ \\
$ |(a_{F}^{(5)})^{JTJ}_{\mu e}|, |(b_{F}^{(5)})^{JTJ}_{\mu e}|	$&$	< 6	$&$	< 1	$&$	< 0.2	$ \\
$ |(a_{F}^{(5)})^{JTK}_{\mu e}|, |(b_{F}^{(5)})^{JTK}_{\mu e}|	$&$	< 8	$&$	< 1	$&$	< 0.2	$ \\ 
$ |(a_{F}^{(5)})^{JTZ}_{\mu e}|, |(b_{F}^{(5)})^{JTZ}_{\mu e}|	$&$	< 8	$&$	< 0.9	$&$	< 0.2	$ \\
$ |(a_{F}^{(5)})^{ZTJ}_{\mu e}|, |(b_{F}^{(5)})^{ZTJ}_{\mu e}|	$&$	< 8	$&$	< 1	$&$	< 0.2	$ \\
$ |(a_{F}^{(5)})^{ZTZ}_{\mu e}|, |(b_{F}^{(5)})^{ZTZ}_{\mu e}|	$&$	< 7	$&$	< 0.9	$&$	< 0.2	$ \\ [4pt]
$ |(H_{F}^{(5)})^{TJTJ}_{\mu e}|, |(H_{F}^{(5)})^{JZTK}_{\mu e}|	$&$	< 7	$&$	< 1	$&$	< 0.2	$ \\
$ |(H_{F}^{(5)})^{TJTK}_{\mu e}|, |(H_{F}^{(5)})^{JZTJ}_{\mu e}|	$&$	< 6	$&$	< 1	$&$	< 0.2	$ \\
$ |(H_{F}^{(5)})^{TJTZ}_{\mu e}|, |(H_{F}^{(5)})^{JZTZ}_{\mu e}|	$&$	< 6	$&$	< 0.9	$&$	< 0.2	$ \\ 
$ |(H_{F}^{(5)})^{TZTJ}_{\mu e}|, |(H_{F}^{(5)})^{XYTJ}_{\mu e}|	$&$	< 6	$&$	< 1	$&$	< 0.2	$ \\
$ |(H_{F}^{(5)})^{TZTZ}_{\mu e}|, |(H_{F}^{(5)})^{XYTZ}_{\mu e}|	$&$	< 7	$&$	< 0.9	$&$	< 0.2	$ \\[4pt]
\hline\hline 
\end{tabular}
\caption{Constraints on $d=5$ SME coefficients 
in units of $10^{-12}$ GeV$^{-1}$.
The results for SINDRUM~II are constraints at 90\% CL. 
The results for Mu2e and COMET are projected constraints 
based on expected rates.}
\label{resultsd5}
\end{table}
Each result is a constraint at 90\% CL 
on an coefficient component in the SCF,
where the indices $J$ and $K\neq J$ are $X$ or $Y$.
The SINDRUM~II result implies sensitivities 
at parts in $10^{12}$ 
to 96 real components.
The constraints involving the coefficients 
$(a_{F}^{(5)})^{\lambda\alpha\beta}_{\mu e}$ and
$(b_{F}^{(5)})^{\lambda\alpha\beta}_{\mu e}$
also represent limits on CPT violation. 

Following a similar approach for the $d=6$ quark-lepton conversion rate~\eqref{d6rate} results in two radial integrals 
\begin{equation}
I_3 = \int \,dr~r^2 \rho^{(p)}(r) f_e^-f_\mu^-,
\quad
I_4 = \int \,dr~r^2 \rho^{(p)}(r) g_e^-g_\mu^-.
\label{integrals2}
\end{equation}
The numerical values of 
$\widetilde{I}_3 = I_3/m_\mu^{5/2}$ and $\widetilde{I}_4 = I_4/m_\mu^{5/2}$
are given in Table~\ref{exptprops2}.
\renewcommand\arraystretch{0.8}
\begin{table}
\centering
\setlength{\tabcolsep}{3pt}
\begin{tabular}{c c c c c}
\hline\hline								
Coefficients	&	SINDRUM~II	&	COMET &	Mu2e	\\	
\hline								
$ |(k_{SV}^{(6)})^T_{uue\mu}|, |(k_{SA}^{(6)})^T_{uue\mu}|	$&$	<6	$&$	<1	$&$	<0.2	$ \\
$ |(k_{SV}^{(6)})^J_{uue\mu}|, |(k_{SA}^{(6)})^J_{uue\mu}|	$&$	<7	$&$	<1	$&$	<0.2	$ \\
$ |(k_{SV}^{(6)})^Z_{uue\mu}|, |(k_{SA}^{(6)})^Z_{uue\mu}|	$&$	<7	$&$	<1	$&$	<0.2	$ \\
$ |(k_{SV}^{(6)})^T_{dde\mu}|, |(k_{SA}^{(6)})^T_{dde\mu}|	$&$	<6	$&$	<1	$&$	<0.2	$ \\
$ |(k_{SV}^{(6)})^J_{dde\mu}|, |(k_{SA}^{(6)})^J_{dde\mu}|	$&$	<7	$&$	<1	$&$	<0.2	$ \\
$ |(k_{SV}^{(6)})^Z_{dde\mu}|, |(k_{SA}^{(6)})^Z_{dde\mu}|	$&$	<7	$&$	<1	$&$	<0.2	$ \\ 
$ |(k_{SV}^{(6)})^T_{sse\mu}|, |(k_{SA}^{(6)})^T_{sse\mu}|	$&$	<10	$&$	<2	$&$	<0.4	$ \\
$ |(k_{SV}^{(6)})^J_{sse\mu}|, |(k_{SA}^{(6)})^J_{sse\mu}|	$&$	<15	$&$	<2	$&$	<0.4	$ \\
$ |(k_{SV}^{(6)})^Z_{sse\mu}|, |(k_{SA}^{(6)})^Z_{sse\mu}|	$&$	<15	$&$	<2	$&$	<0.4	$ \\
$ |(k_{VS}^{(6)})^{T}_{uu e\mu}|, |(k_{VP}^{(6)})^{T}_{uu e\mu}|	$&$	<30	$&$	<4	$&$	<0.8	$ \\
$ |(k_{VS}^{(6)})^{T}_{dde\mu}|, |(k_{VP}^{(6)})^{T}_{dde\mu}|	$&$	<30	$&$	<4	$&$	<0.7	$ \\ [4pt]
$|(k_{ST}^{(6)})^{TJ}_{uue\mu}|, |(k_{ST}^{(6)})^{JZ}_{uue\mu}|	$&$	<7	$&$	<1	$&$	<0.2	$ \\ 
$|(k_{ST}^{(6)})^{TZ}_{uue\mu}|, |(k_{ST}^{(6)})^{XY}_{uue\mu}|	$&$	<7	$&$	<1	$&$	<0.2	$ \\
$|(k_{ST}^{(6)})^{TJ}_{dde\mu}|, |(k_{ST}^{(6)})^{JZ}_{dde\mu}|	$&$	<7	$&$	<1	$&$	<0.2	$ \\ 
$|(k_{ST}^{(6)})^{TZ}_{dde\mu}|, |(k_{ST}^{(6)})^{XY}_{dde\mu}|	$&$	<7	$&$	<1	$&$	<0.2	$ \\ 
$|(k_{ST}^{(6)})^{TJ}_{sse\mu}|, |(k_{ST}^{(6)})^{JZ}_{sse\mu}|	$&$	<15	$&$	<2	$&$	<0.4	$ \\ 
$|(k_{ST}^{(6)})^{TZ}_{sse\mu}|, |(k_{ST}^{(6)})^{XY}_{sse\mu}|	$&$	<15	$&$	<2	$&$	<0.4	$ \\ 
$|(k_{VV}^{(6)})^{TT}_{uue\mu}|, |(k_{VA}^{(6)})^{TT}_{uue\mu}|	$&$	<20	$&$	<4	$&$	<0.7	$ \\ 
$|(k_{VV}^{(6)})^{TJ}_{uue\mu}|, |(k_{VA}^{(6)})^{TJ}_{uue\mu}|	$&$	<25	$&$	<4	$&$	<0.7	$ \\ 
$|(k_{VV}^{(6)})^{TZ}_{uue\mu}|, |(k_{VA}^{(6)})^{TZ}_{uue\mu}|	$&$	<25	$&$	<4	$&$	<0.7	$ \\ 
$|(k_{VV}^{(6)})^{TT}_{dde\mu}|, |(k_{VA}^{(6)})^{TT}_{dde\mu}|	$&$	<20	$&$	<4	$&$	<0.7	$ \\ 
$|(k_{VV}^{(6)})^{TJ}_{dde\mu}|, |(k_{VA}^{(6)})^{TJ}_{dde\mu}|	$&$	<20	$&$	<4	$&$	<0.7	$ \\ 
$|(k_{VV}^{(6)})^{TZ}_{dde\mu}|, |(k_{VA}^{(6)})^{TZ}_{dde\mu}|	$&$	<20	$&$	<4	$&$	<0.7	$ \\ [4pt]
$|(k_{VT}^{(6)})^{TTJ}_{uue\mu}|, |(k_{VT}^{(6)})^{TJZ}_{uue\mu}|	$&$	<25	$&$	<4	$&$	<0.7	$ \\ 
$|(k_{VT}^{(6)})^{TTZ}_{uue\mu}|, |(k_{VT}^{(6)})^{TXY}_{uue\mu}|	$&$	<25	$&$	<4	$&$	<0.7	$ \\ 
$|(k_{VT}^{(6)})^{TTJ}_{dde\mu}|, |(k_{VT}^{(6)})^{TJZ}_{dde\mu}|	$&$	<20	$&$	<4	$&$	<0.7	$ \\ 
$|(k_{VT}^{(6)})^{TTZ}_{dde\mu}|, |(k_{VT}^{(6)})^{TXY}_{dde\mu}|	$&$	<20	$&$	<4	$&$	<0.7	$ \\ [4pt]
\hline\hline
\end{tabular}
\caption{Constraints on $d=6$ SME coefficients 
in units of $10^{-13}$ GeV$^{-2}$.
The results for SINDRUM~II are constraints at 90\% CL. 
The results for Mu2e and COMET are projected constraints 
based on expected rates.}
\label{resultsd6}
\end{table}
Again interpreting the SINDRUM~II result as a constraint
on the time-averaged conversion rate $\overline{\omega}_{\rm conv}$,
we establish and present in Table~\ref{resultsd6} first constraints on each 
of the $d=6$ coefficient components in the SCF,a
where $J$ is $X$ or $Y$.
The results show that SINDRUM~II 
probes 148 real coefficient components
arising from 4-point CLFC quark-lepton interactions to parts in $10^{13}$.
The constraints for components with one or three spacetime indices
also represent constraints on CPT violation. 
Table~\ref{resultsd6} also displays the corresponding projections 
for the Mu2e and COMET experiments.

Together with the enhanced resolving power of a sidereal analysis,
we suggest that conversion experiments at Fermilab and J-PARC could ultimately yield orders-of-magnitude improvements over the present results. This provides an optimistic outlook for future tests of Lorentz and CPT violation in CLFC processes.

\section*{Acknowledgments}
This work was supported by the Deutsche Forschungsgemeinschaft 
under the Heinz Maier Leibnitz Prize BeyondSM HML-537662082,
and by the Indiana University Center for Spacetime Symmetries.

\end{document}